\documentclass{PoS}

\title{Neutron Stars and the EOS}

\ShortTitle{Neutron Stars and the EOS}

\author{\speaker{Madappa Prakash}%
         \thanks{MP thanks his collaborators Mark Alford, Sophia Han, James Lattimer and Sergey Postnikov with great pleasure.
          Research support from the US Department of Energy under contract \#DE-FG02-93ER-40756 is gratefully acknowledged.
 }\\
        Department of Physics and Astronomy, Ohio University, Athens, OH, 45701 \\
        E-mail: \email{prakash@harsha.phy.ohiou.edu}}

\abstract{Implications of recently well-measured neutron star masses, particularly near and above 2 solar masses, for the equation of state (EOS) of neutron star matter are highlighted.  Model-independent upper limits to thermodynamic properties in neutron stars, which only depend on the neutron star maximum mass, established from causality considerations are presented. The need for non-perturbative treatments of quark matter in neutron stars is stressed through studies of self-bound  quark matter stars, and of nucleon-quark hybrid stars.  
The extent to which several well-measured masses and radii of individual neutron stars can establish a model-independent EOS through an inversion of the stellar structure equations is briefly discussed.}

\FullConference{8th International Workshop on Critical Point and Onset of 
Deconfinement\\
		 March 11-15, 2013\\
		 Napa, California, USA}
		 
\def\simge{\mathrel{\rlap{\raise 0.511ex
       \hbox{$>$}}{\lower 0.511ex \hbox{$\sim$}}}}
\def\simle{\mathrel{\rlap{\raise 0.511ex
        \hbox{$<$}}{\lower 0.511ex \hbox{$\sim$}}}}

\newcommand{\ee}{\end{eqnarray}} 
\newcommand{\be}{\begin{eqnarray}} 
\newcommand{\ec}{\end{center}} 
\newcommand{\bc}{\begin{center}} 
\newcommand{\eea}{\end{eqnarray}} 
\newcommand{\bea}{\begin{eqnarray}} 
\newcommand{\bd}{\begin{description}} 
\newcommand{\ed}{\end{description}} 
\newcommand{\bi}{\begin{itemize}} 
\newcommand{\ei}{\end{itemize}}

\newcommand{\bis}{\begin{itemstep}} 
\newcommand{\eis}{\end{itemstep}}

\begin{document}

\section{Introduction}
The two most basic properties of a neutron star are its mass $M$ and radius $R$. The importance of these physical traits is highlighted by several other observables including~\cite{LP:07} 
\begin{enumerate}
\item The binding energy B.E. of a neutron star: 
\be
B.E \simeq  (0.6\pm0.05)~\frac{GM^2}{Rc^2}~\left(1-\frac{GM}{Rc^2}\right)^{-1} \,.
\ee
Nearly 99\% this B.E. is carried by neutrinos emitted during the birth of a neutron star. 
\item Minimum spin periods of rotation:
\be
P_{min}(M_{max}) &=& 0.83~\left(\frac {M_{max}}{\rm M_\odot}\right)^{-1/2} 
\left(\frac{R_{max}}{10~{\rm km}}\right)^{3/2} ~{\rm ms} \,, \nonumber \\
P_{min}(M)&\simeq&(0.96\pm 0.3) ~\left(\frac {M}{\rm M_\odot}\right)^{-1/2} 
\left(\frac{R}{10~{\rm km}}\right)^{3/2} ~{\rm ms} \,, 
\ee
where $M_{max}$ and $R_{max}$ refer to the non-rotating maximum mass spherical configurations, and the second relation refers to an arbitrary mass not too close to the maximum mass.
\item Moment of Inertia: \\
\be
I_{max} = 0.6\times 10^{45} 
\frac {(M_{max}/{\rm M_\odot}) (R_{max}/10~{\rm km})^2}
{1-0.295(M_{max}/{\rm M_\odot})/(R_{max}/10~{\rm km}) }
~~{\rm g~cm^2} \,.
\ee
Accurate pulse timing techniques are needed to measure $I$ in a double neutron star binary.
\end{enumerate}  
For a list of other observables that are significantly influenced by $M$ and $R$, see Ref. \cite{LP:07}.  Through the general relativistic (TOV) equations of stellar structure \cite{TOV}, $M,~R,~B.E.~,I$, and the surface red-shift $\phi_s = (1-2GM/Rc^2)^{-1/2}-1$ can be calculated once the equation of state (EOS) of neutron star matter (the relationship between pressure $p$ and energy density $\epsilon$ at every location in the star) is provided. The one-to-one correspondence between the EOS ($p~{\rm vs}~\epsilon$) and the observed $M~{\rm vs}~R$ curve can be used to advantage to model-independently determine the EOS of neutron star matter as will be discussed later.  

\section{Neutron star masses}
\begin{figure}[thb]
\vskip -2cm
\begin{center}
{\includegraphics[width=380pt,angle=0]{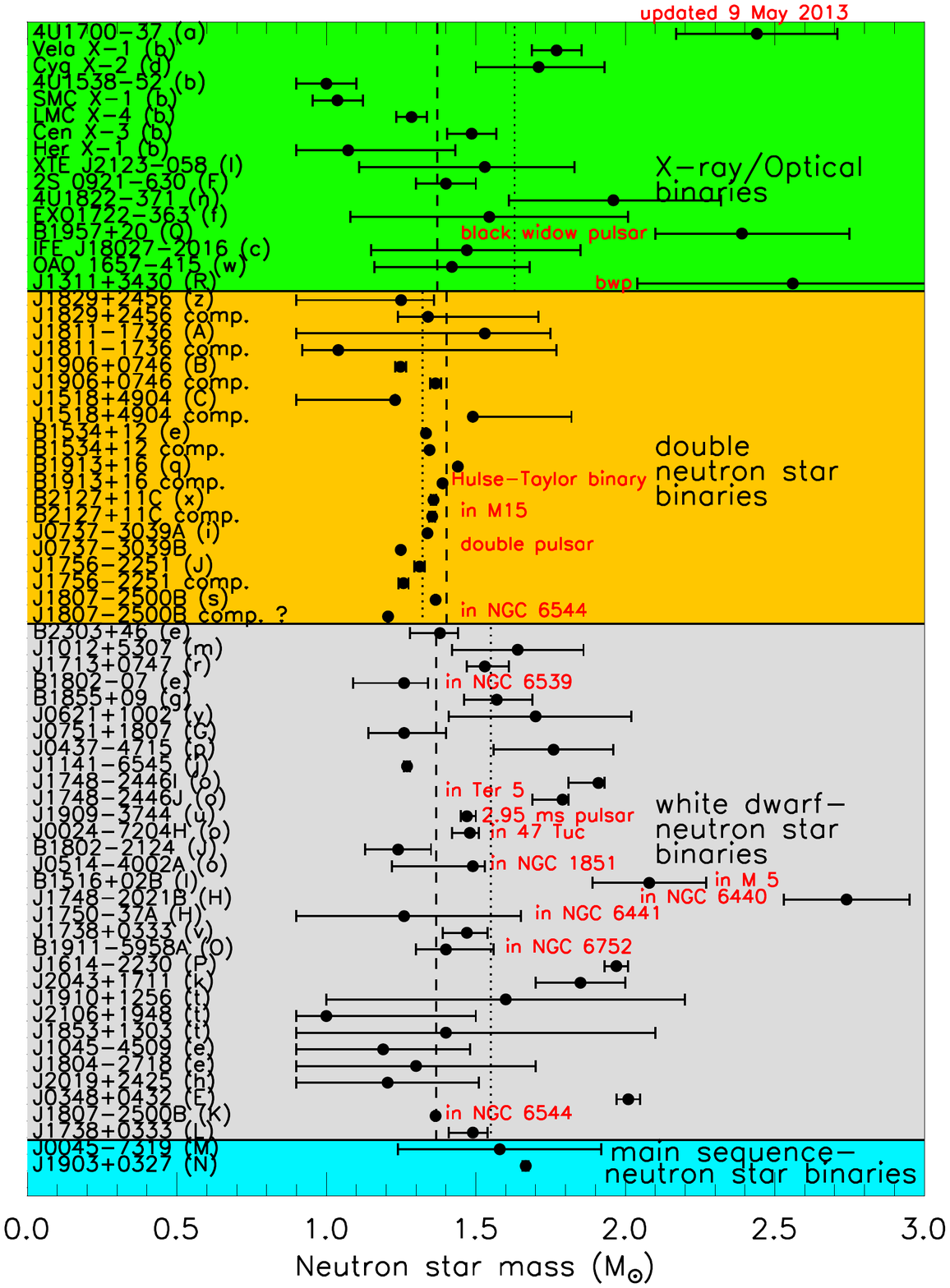}}
\end{center}
\vskip -2.0cm
\caption{Measured neutron star masses with 1-$\sigma$ errors from the compilation of Lattimer, who maintains a contemporary table, figure and references in  http://www.stellarcollapse.org. The vertical dahed (dotted) line shows the mean (error weighted mean) mass in each category. }
\label{masses}
\end{figure}

Figure \ref{masses} shows the measured  neutron star masses from the recent compilation of Lattimer. The most accurate measurements are for pulsars in bound binary systems in which five Keplerian parameters can be precisely measured by pulse timing techniques \cite{manchester1977}, including (i) the binary period $P$, (ii) the projection
of the pulsar's semimajor axis on the line of sight $a_p\sin i$ (where
$i$ is the binary inclination angle), (iii) the eccentricity $e$, and (iv and v)  the
time and longitude of periastron $T_0$ and $\omega$.  Binary pulsars are compact systems and general
relativistic effects can often be observed.  These include (a) the advance
of the periastron of the orbit $\dot\omega$, (b) the
combined effect of variations in the tranverse Doppler shift and
gravitational redshift around an elliptical orbit $\gamma$, (c) the orbital period decay due to the emission of gravitational
radiation $\dot P$, and (d) Shapiro
time delay, $\delta_S$, which is caused by the propagation of the pulsar signal
through the gravitational field of its companion.  The happenstance that all these observables are different functions of the individual masses of the pulsar and its companion facilitates a precise determination of both masses in this over-determined system when all of the above quantities can be measured with precision (for recent pedagogical accounts, see, e.g. Refs. \cite{LP11,L12}). 
However, only in a handful of cases have measurements of two or more general relativistic effects been possible. 
The extent to which measurements have yielded  neutron star masses in observations of X-ray/Optical, double neutron star, white-dwarf-neutron star and main sequence-neutron star binaries is highlighted in Figure \ref{masses}.

%%%%%%%%%%%%%%%%%%%%%%%%%%%%%%%%%%%%%%%%
%
%
%%%%%%%%%%%%%%%%%%%%%%%%%%%%%%%%%%%%%%%%

\section{Implications of 2 ${\rm M}_\odot$ neutron stars}
Recently, the outstanding discoveries of the $1.97\pm 0.04~{\rm M}_\odot$ pulsar in PSR J1614-2230 \cite{demorest10} and  
 $2.01\pm 0.04~{\rm M}_\odot$ pulsar in PSR J0348+0432 \cite{antoniadis13} have caused much excitement inasmuch as these well-measured masses severely delimit the EOS of neutron star matter.   Masses well in excess of 2 ${\rm M}_\odot$, 
 albeit with large uncertainties,  have also been reported, as e.g., $2.44^{+0.27}_{-0.27}~{\rm M}_\odot$ for 4U 1700-377, $2.39^{+0.36}_{-0.29}~{\rm M}_\odot$ for PSR B1957+20 both in X-ray binaries, and $2.74\pm 0.21~{\rm M}_\odot$ for 
 J1748-2021B in neutron star-white dwarf binaries. From the perspective of the EOS of strongly interacting matter, these developments have allowed us to establish firm upper limits to the energy density $\epsilon$, pressure $p$, baron number density $n$, and chemical potential $\mu$ in neutron star matter as summarized below.
 
 \subsection*{Maximally compact EOS}
 The most compact and massive configurations occur when the low-density EOS is ``soft'' and the high-density EOS is ``stiff'' \cite{Haensel89,Koranda97}. 
 Using limiting forms in both cases, the maximally compact EOS is therefore given by the pressure ($p$) versus energy density $(\epsilon)$ relation
 \be
 p = 0 \quad {\rm for} \quad \epsilon < \epsilon_0\,; \quad p = \epsilon - \epsilon_0 \quad {\rm for} \quad \epsilon > \epsilon_0\,,
 \ee
 the stiff EOS being at the causal limit as $dp/d\epsilon=(c_s/c)^2=1$, where $c_s$ is the adiabatic speed of sound. This EOS has a single parameter $\epsilon_0$ and the structure (TOV) equations scale with it according to \cite{Witten84}
 \be
 \epsilon \propto \epsilon_0\,, \qquad p \propto \epsilon_0\,, \qquad m \propto \epsilon_0^{-1/2}, \qquad {\rm and} \qquad 
 r \propto \epsilon_0^{-1/2} \,,
 \ee
 where $m$ is the star's enclosed mass and $r$ its radius.
 Utilizing these properties, the mass and radius of the configuration that maximizes the compactness ratio $(GM_{max}/R_{max}c^2)$ were found to be \cite{Koranda97,LP11}
 \be
 M_{max} = 4.09~ (\epsilon_s/\epsilon_0)^{1/2} ~{\rm M}_\odot\,, \quad 
 R_{max} = 17.07~ (\epsilon_s/\epsilon_0)^{1/2}~{\rm km}\,,  \quad {\rm and} \quad
 BE_{max} = 0.34~M_{max}c^2  
 \label{maxMR}
 \ee  
 where $\epsilon_s \simeq 150~{\rm MeV~fm}^{-3}$ is the energy density at the nuclear saturation density of $n_0=0.16~{\rm fm}^{-3}$. If the EOS is deemed known up to $\epsilon_0\sim 2\epsilon_s$, the maximally compact EOS yields $M_{max} \sim 3~{\rm M}_\odot$.    
 
 The upper limits on the corresponding thermodynamic variables are \cite{Koranda97,LP11}:
 \be
 \epsilon_{max} = 3.034~\epsilon_0\,, \quad p_{max} = 2.034~\epsilon_0\,, \quad \mu_{max} = 2.251~\mu_0\,, 
 \quad {\rm and} \quad n_{max} = 2.251~(\epsilon_0/\mu_0)\,,
\label{maxthermo}
\ee
 where $\mu_0\simeq 930$ MeV is the mass-energy of iron nuclei per baryon in a star with a normal crust. Combining Eqs. 
 (\ref{maxMR}) and (\ref{maxthermo}), we arrive at the result \cite{LP11}
 \be
 \epsilon_{max} \leq 50.8~ \epsilon_s ~({\rm M}_\odot/M_{max})^2\,, 
 \label{Ultimate}
 \ee
 a relation that enables us to appreciate the impact of the maximum mass of a neutron star on the ultimate energy density of cold observable matter.  If the largest measured mass represents the true neutron star maximum mass, it sets upper limits on the central energy density, pressure, baryon number density and chemical potential. In the case of the 1.97 M$_\odot$, these limits turn out to be
 \be
 \epsilon_{max} < 1.97~{\rm GeV~fm}^{-3}, \quad p_{max} < 1.32~{\rm GeV~fm^{-3}}, \quad 
 n_{max} < 1.56~{\rm fm^{-3}}, \quad \mu_{max}  < 2.1~{\rm GeV}\,.  
 \ee
 %
%%%%%%%%%%%%%%%%%%%%%%%%%%%%%%%%%%%%%%%%
%
\begin{figure}[thb]
\vskip -1cm
\begin{center}
{\includegraphics[width=300pt,angle=90]{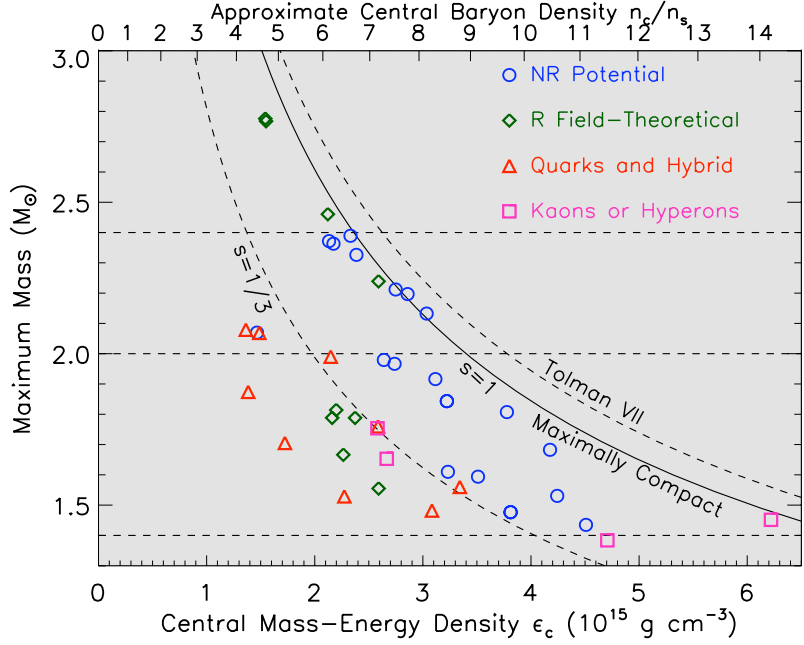}}
\end{center}
\vskip -1.5cm
\caption{Maximum mass versus central mass-energy density (bottom $x$-axis) and central baryon density (top $x$-axis) 
according to Eq. (3.5) labelled $s=1$. 
The curve labelled $s=1/3$ corresponds to  $p=(\epsilon-\epsilon_0)/3$ characteristic of commonly used quark matter EOSs.  Results for the Tolman VII solution [11]  with  
$\epsilon=\epsilon_c(1-(r/R)^2)$ and for various model calculations of neutron star matter - see inset for legends - are as shown. Figure adapted from Ref. [12].
}
\label{ultimate}
\end{figure}
%
%%%%%%%%%%%%%%%%%%%%%%%%%%%%%%%%%%%%%%%%
%
%\cite{Tolman39} 
%Ref. \cite{LP05}.}
Figure \ref{ultimate} depicts the relation between the maximum mass and the central energy density  $\epsilon_c$ and 
baryon number density $n_c$ for the maximally compact EOS (also labelled $s=1$). Note the substantial reductions in the energy density and baryon number density if the observed mass is 2.4 M$_\odot$.  Clearly, accurate observations of masses well in excess of 2 M$_\odot$ have profound implications for the EOS of neutron star matter.

The analysis presented above was performed for the general EOS $p=s(\epsilon-\epsilon_0)$ in Ref. \cite{LP05} for various values of $s$. The case $s=1/3$ and $\epsilon_0=4B$ corresponds the MIT bag model EOS with $B$ being the bag constant.  
Maximally compact configurations in this case are characterized by
 \be
 M_{max} &=& 2.48~ (\epsilon_s/\epsilon_0)^{1/2} ~{\rm M}_\odot \quad 
 R_{max} = 13.56~ (\epsilon_s/\epsilon_0)^{1/2}~{\rm km}\,,  \quad {\rm and} \quad 
 BE_{max} = 0.21~M_{max}c^2
 \nonumber \\
 \label{maxMRQM}
\epsilon_{max} &\simeq& 30\left(\frac{{\rm M_\odot}}{M_{max}}\right)^2~ \epsilon_s\, , 
\quad p_{max}  \simeq 7.9\left(\frac{{\rm M_\odot}}{M_{max}}\right)^2~ \epsilon_s\, , 
 \quad n_{B,max}   \simeq 27\left(\frac{{\rm M_\odot}}{M_{max}}\right)^2~ n_s\, , 
  \quad {\rm and} \nonumber \\
 \quad \mu_{B,max} &\simeq& 1.46~{\rm GeV}\,,  
\label{maxthermoQM}
 \ee  
where a value of $\mu_0=930$ MeV was used as self-bound quark stars are expected to have a very thin crust (that does not affect $M$ and $R$ significantly) of normal matter. The $M_{max}$ versus $\epsilon_c$ curve for $s=1/3$ shown in Fig. \ref{ultimate} lies a factor of $\sim 0.6$ below the $s=1$ curve. It was verified in Ref. \cite{LP05} that effects of adding QCD corrections, finite strange quark mass and CFL gaps makes the EOS more attractive and less compact. Also noteworthy is the relatively low value of the baryon chemical potential (1.46 GeV), which calls for non-perturbative treatments of quark matter.

\subsection*{Hybrid stars}
\vskip -0.25cm
The two solar mass measurements also raise questions about limits to the extent of hyperons, Bose (e.g. kaon) condensates, quarks, etc., in the cores of neutron stars. Here the findings of a recent study \cite{Alford13} that examined hybrid stars, assuming a single first-order phase transition between nuclear and quark matter, with a sharp interface between the quark matter core and nuclear matter mantle, is  summarized.  To establish generic conditions for stable hybrid stars, the EOS of dense matter was approximated by 

\be
\epsilon(p) = \left\{\!
\begin{array}{ll}
\epsilon_{\rm NM}(p) & \quad p<p_{trans}\\
\epsilon_{\rm NM}(p_{trans})+\Delta\epsilon+c_{\rm QM}^{-2} (p-p_{trans}) & \quad p>p_{trans}
\end{array}
\right.
\label{eqn:EoSqm1}
\ee
where $\epsilon_{\rm NM}(p)$ is the nuclear matter equation of state, $\Delta\epsilon$ is the discontinuity in energy density $\epsilon$ at the transition pressure $p_{trans}$, and $c_{QM}^2$ is the squared speed of sound of quark matter taken to be constant with density (as in a classical ideal gas) but varied in the range 1/3 (roughly characteristic of perturbative quark matter) to 1 (causal limit).   For illustrative purposes, we use two examples for $\epsilon_{{\rm NM}}(p)$: a 
relativistic mean field model labelled NL3 \cite{Shen:2011kr} 
and a non-relativistic potential model labelled HLPS,
corresponding to ``EoS1'' in Ref.~\cite{Hebeler:2010jx}.  Insofar as HLPS is softer than NL3, these EOSs provide a contrast  at low density. Figure 4 explores maximum mass contours as a function of the quark matter EOS parameters for HLPS and NL3 choices of  the nuclear matter EOS.

%%%%%%%%%%%%%%%%%%%%%%%%%%%%%%%%%%%%%%%%
%
\begin{figure}[thb]
\vskip -0.25cm
%\begin{center}
{\includegraphics[width=220pt,angle=0]{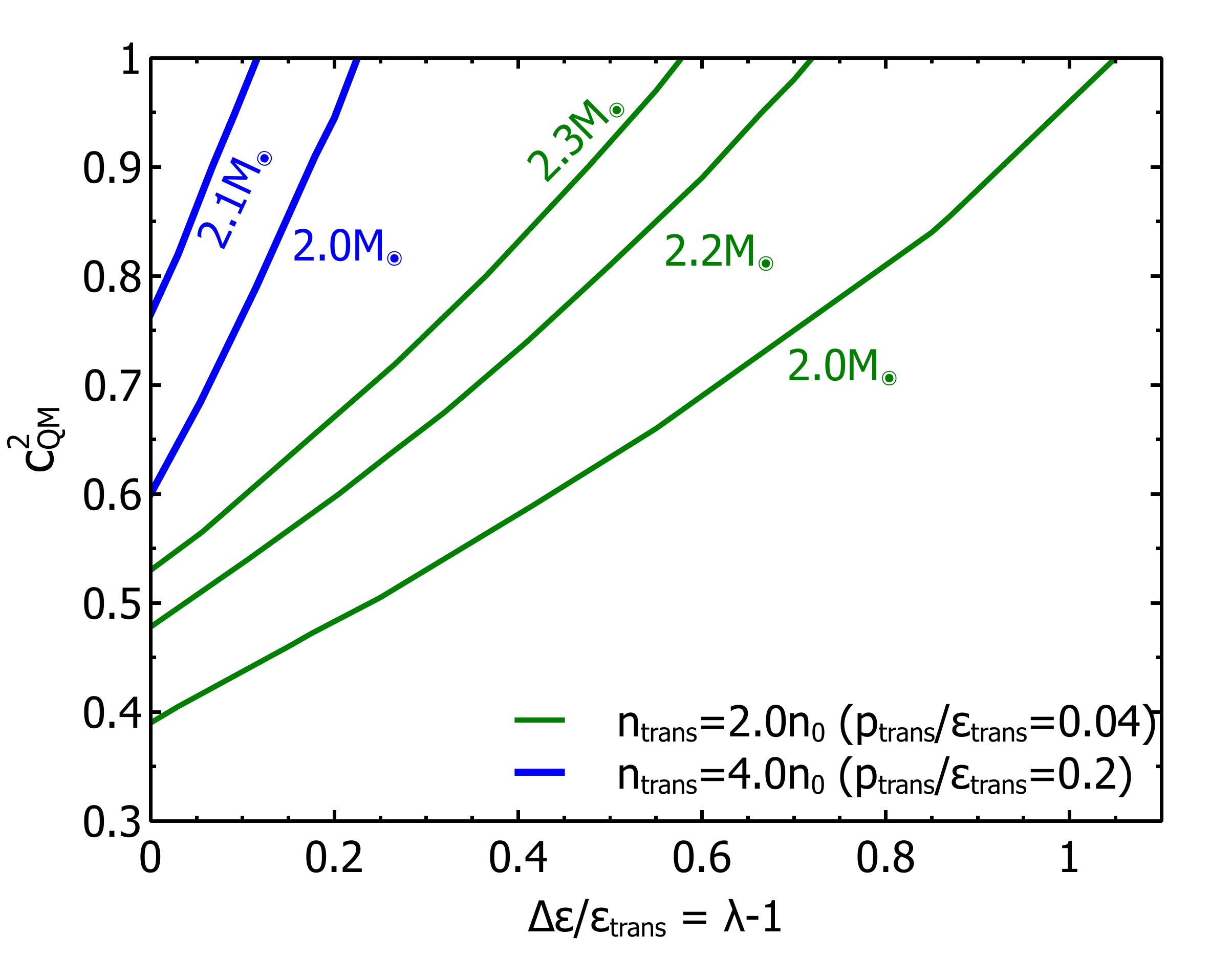}}
{\includegraphics[width=220pt,angle=0]{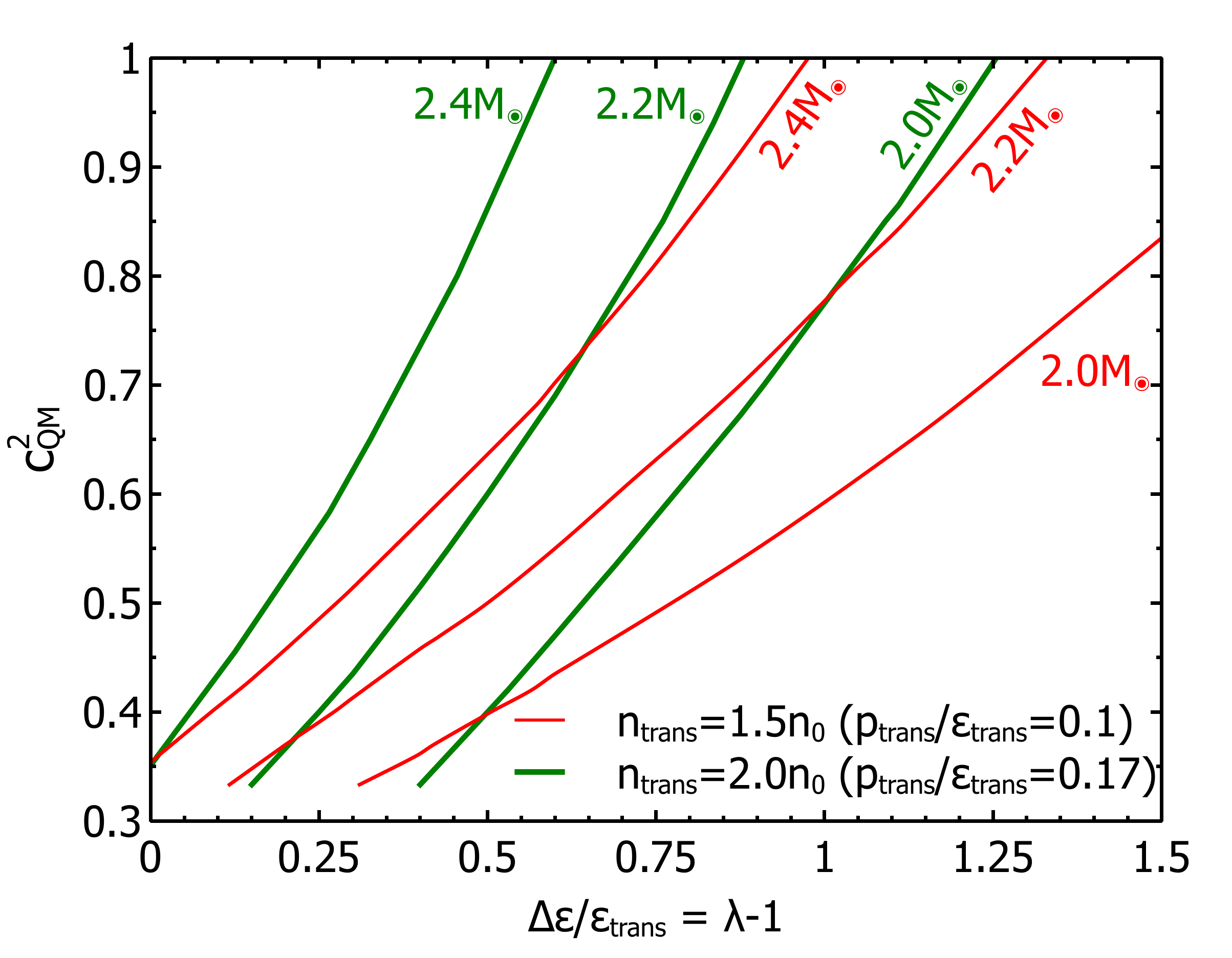}}
%\end{center}
%\vskip -2.0cm
\caption{Contour plot of the mass of the heaviest hybrid star as a function of quark matter EOS parameters $p_{trans}/\epsilon_{trans}$, $c_{QM}^2$, and $\Delta\epsilon/\epsilon_{trans}$ (a shifted version of $\lambda$ in Ref. [16] 
%\cite{Schaeffer83}) 
for HLPS (left panel) and NL3 (right panel) nuclear matter.The thin (red), medium (green) and thick (blue) lines are for nuclear to quark transition at 
$n_{trans}=1.5n_0,~2n_0$ and $4n_0$, respectively. Figure adapted from Ref. [13].}
%\cite{Alford13}.}
\label{maxmasses}
\end{figure}
%
%%%%%%%%%%%%%%%%%%%%%%%%%%%%%%%%%%%%%%%%

A principal finding of this work was  that  it is possible to get heavy hybrid stars in excess of 2 M$_\odot$ for reasonable parameters of the quark matter EOS. It requires not-too-high transition density ($n\sim 2n_0$), low enough energy density discontinuity $\Delta\epsilon < 0.5~\epsilon_{trans}$, and high enough spped of sound $c_{QM}^2 \geq 0.4$. It is worthwhile to note that free quark matter is characterized by $c_{QM}^2=1/3$, and a value of $c_{QM}^2$ well above 1/3 is an indication that quark matter is strongly coupled. Clearly, non-perturbative treatments of quark matter are indicated. 

\section{Toward a model-independent EOS of neutron star matter}
\vskip-0.25cm
%%%%%%%%%%%%%%%%%%%%%%%%%%%%%%%%%%%%%%%%
%
\begin{figure}[thb]
\vskip -0.25cm
{\includegraphics[width=220pt,angle=0]{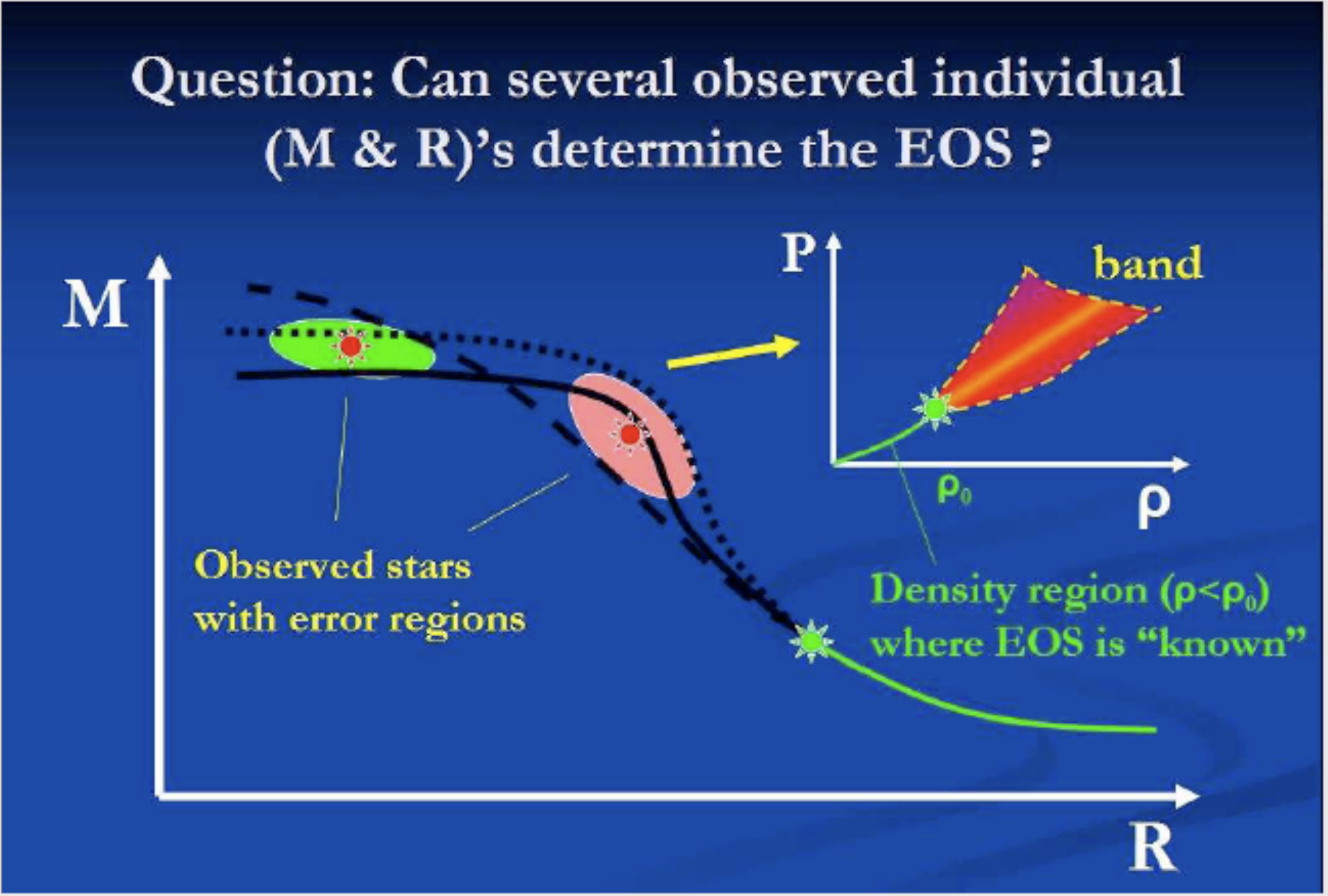}}
{\includegraphics[width=220pt,height=150pt,angle=0]{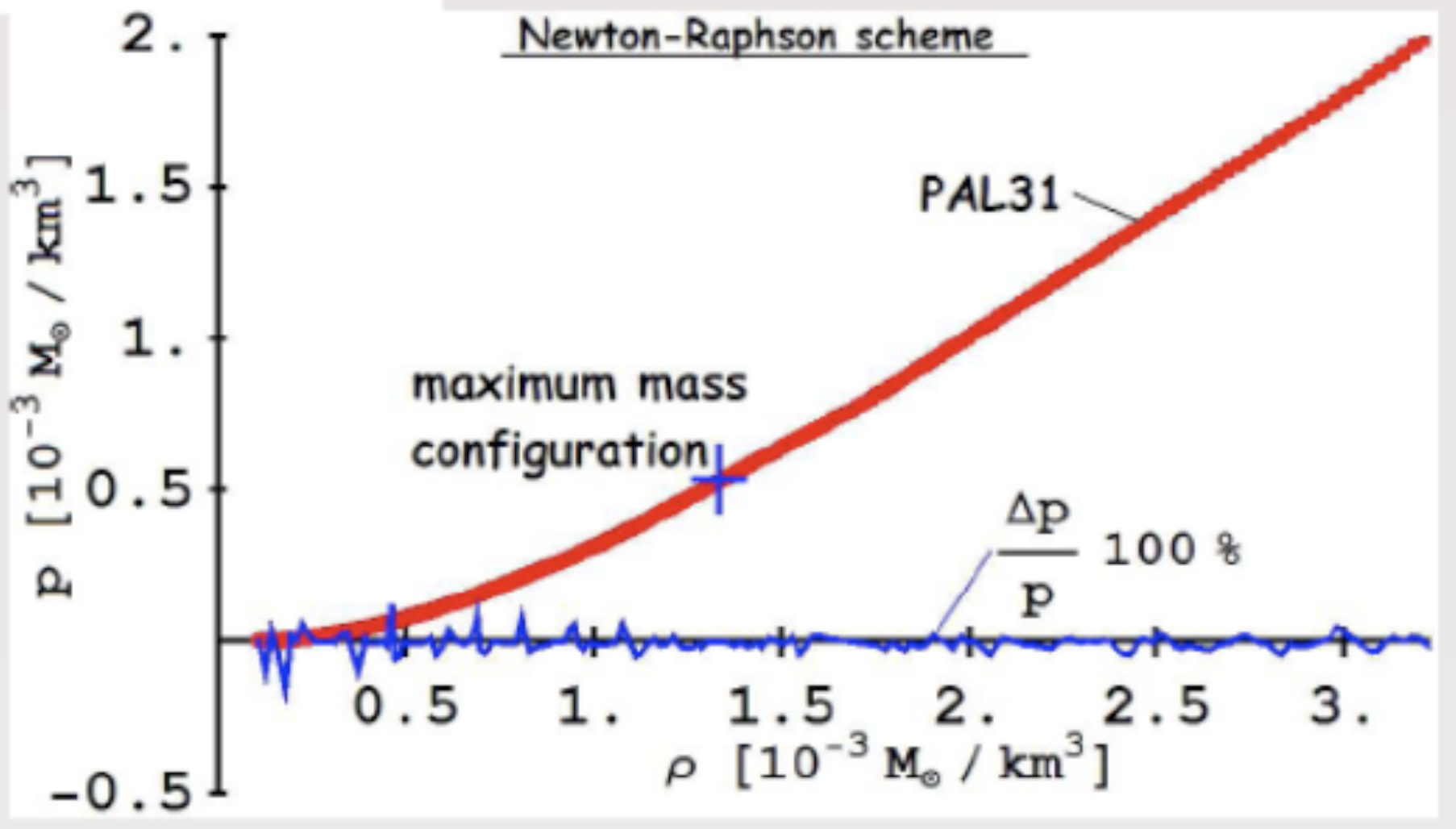}}
\caption{ Deconstructing a neutron star with a physically motivated nucleonic EOS. Left panel figure courtesy Postnikov. Right panel figure adapted from Ref. [17].} 
%\cite{Prakash09}.}
\label{deconstruct}
\end{figure}
%
%%%%%%%%%%%%%%%%%%%%%%%%%%%%%%%%%%%%%%%%
Accurately measured masses and radii of several individual neutron stars can uniquely determine the dense matter EOS in a model-independent manner. The method, first outlined by Lindblom \cite{Lindblom92}, exploits the one-to-one correspondence between an EOS and the $M-R$ curve generated through the use of the (TOV) equations of stellar structure \cite{TOV} which can be rewritten as \cite{Prakash09,Postnikov10}
\be
\frac {dr^2}{dh} = -2r^2 \frac {r-2m}{m+4\pi r^3P} \qquad {\rm and} \qquad \frac {dm}{dh} = -4\pi r^3\rho \frac {r-2m}{m+4\pi r^3P}\,,
\label{rtov}
\ee
where the pressure $p(h)$ and mas-energy density $\rho(h)$, which serve as input, contain the EOS. Above, the variable $h$ is defined through $dh=dp/(p+\rho(p))$. The advantages of this reformulation are that the enclosed mass $m$ and radius $r$ are now dependent (on $h$ and thus the EOS) variables, and $h$ is finite both at the center and surface of the star. The deconstruction procedure begins with a known EOS up to a certain density, taking small increments in mass and radius, and adopting an iterative scheme based on Eqs. (\ref{rtov}) to reach the new known mass and radius. Alternatively, one can solve Eqs. (\ref{rtov}) from the center to the surface with an assumed form of the EOS using a Newton-Raphson scheme to obtain the known mass and radius. 
The right panel of Fig. \ref{deconstruct} shows the extent to which the EOS is reconstructed (from the latter scheme)
for the case in which the masses and radii are assumed to be those that result from the EOS of PAL31 \cite{PAL}. The two iterative schemes described above both yield results to hundredths of percent accuracy. 

The number of neutron stars for which simultaneous measurements of masses and radii are available, and the inherent errors in measurements will determine the accuracy with which the EOS can be determined. To date, data on radii to the same level of accuracy that radio pulsar measurements on masses of neutron stars have afforded us do not exist. Using the currently available data with somewhat large uncertainties on estimated masses and radii, Steiner et al. \cite{Steiner10} have arrived at  probability distributions for pressure as a function of energy density using the $M-R$ probability distributions through a Bayesian analysis that assumed a parametrized form for the EOS. These developments highlight the need for 
precise measurements of the mass and radius of the same neutron star as it would be a first and an outstanding achievement in neutron star research.  As the theory is already in place, several such measurements offer the promise to pin down the 
EOS of neutron star matter model-independently.

\section{Conclusions}
\vskip -0.25cm
 The largest well measured mass of a neutron star - insofar as it reflects the true maximum mass - establishes the  ultimate energy density of observable cold matter. 
Through the use of a maximally compact EOS, it has been possible 
to establish rigorous upper bounds on the EOS characteristics such as the central energy density $\epsilon_c$, pressure $p_c$ and chemical potential $\mu_c$. If the recently well-measured 2 M$_\odot$ is taken as a proxy for the true maximum mass, the upper bounds on $\epsilon_c$ and $\mu_c$ are $\sim 2~{\rm GeV~fm^{-3}}$ when the high-density EOS is as stiff as causality permits (squared speed of sound $c_s^2=1)$.  Maximally compact quark stars modeled with $c_s^2\simeq1/3$ (characteristic of perturbative QCD) yields  $\epsilon_c\sim 1~{\rm GeV~fm^{-3}}$ and $\mu_c \sim 1.5~{\rm GeV}$ stressing the need for non-perturbative treatments of cold quark matter. A study of nucleon-quark hybrid stars performed with varying $c_s^2$ (but constant with density) for quark matter leads to a similar conclusion inasmuch as $c_s^2 > 0.4$ are required to support 2 M$_\odot$. 
Several well-measured masses and radii of individual neutron stars 
can establish a model-independent EOS through an inversion of the stellar structure equations. 

%%%%%%%%%%%%%%%%%%%%%%%%%%%%%%%%%%%%%%%%%%%%%%%%%%%%%%%%%%%

\end{document}